\documentclass[aps,prl,showkeys,
longbibliography,
reprint,
twocolumn, 
nofootinbib,
floatfix]{revtex4-1}

\usepackage[american]{babel}
\usepackage{amsmath}
\usepackage{amsfonts}
\usepackage{graphicx}
\usepackage{tikz}
\usepackage{subcaption}
\usepackage[colorlinks=true,hyperfootnotes=true,breaklinks=true]{hyperref}
\usepackage{cleveref}

\DeclareMathOperator{\sign}{\mathrm{sign}}

\begin{document}

\title{Expulsion from structurally balanced paradise}

\author{Krzysztof Malarz}
\thanks{\includegraphics[width=10pt]{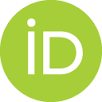}~\href{https://orcid.org/0000-0001-9980-0363}{0000-0001-9980-0363}}
%% \homepage{http://home.agh.edu.pl/malarz/}
\email{malarz@agh.edu.pl}
\affiliation{AGH University of Science and Technology,
Faculty of Physics and Applied Computer Science,
al. Mickiewicza 30, 30-059 Krak\'ow, Poland}

\author{Maciej Wo{\l}oszyn}
\thanks{\includegraphics[width=10pt]{ORCID.png}~\href{http://orcid.org/0000-0001-9896-1018}{0000-0001-9896-1018}}
%% \homepage{http://home.agh.edu.pl/woloszyn/}
\email{woloszyn@agh.edu.pl}
\affiliation{AGH University of Science and Technology,
Faculty of Physics and Applied Computer Science,
al. Mickiewicza 30, 30-059 Krak\'ow, Poland}

\begin{abstract}
We perform simulations of structural balance evolution on a triangular lattice using the heat-bath algorithm.
In contrast to similar approaches---but applied to analysis of complete graphs---the triangular lattice topology successfully prevents the occurrence of even partial Heider balance.
Starting with the state of Heider's paradise, it is just a matter of time when the evolution of the system leads to an unbalanced and disordered state.
The time of the system relaxation does not depend on the system size.
The lack of any signs of a balanced state was not observed in earlier investigated systems dealing with the structural balance.
\end{abstract}
\date{\today}
\maketitle

% \textbf{Social processes modeled on various types of complex and regular networks have been studied intensively in recent years.
% Among them, one of the current issues in social network analysis is the structural balance (or the Heider balance) describing systems of hostile and friendly attitudes represented by negative and positive links among the agents and their time evolution.
% Our results indicate that total vanishing of the ordered and balanced phases of the system is possible in certain topologies, which is very different from the results already known for the case of complete graphs.
% We show that in the case of a triangular lattice, the presence of thermal noise successfully prevents even partial Heider balance, thus leading to disordered and unbalanced states.
% In other words, the critical temperature of the system approaches zero. 
% The demonstrated phenomenon helps to achieve a deeper understanding of interactions that play an important role in social processes.}

%% ###########################################################
\section{Introduction}
%% ###########################################################

\begin{figure}[htbp]
\includegraphics[width=\columnwidth]{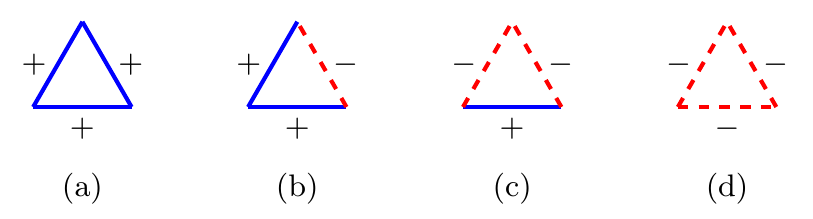}
\caption{\label{fig:triads} Heider's triads corresponding to balanced [(a) and (c)] and imbalanced [(b) and (d)] states. Continuous blue lines and dashed red lines represent friendly and hostile relations, respectively.}
\end{figure}

The structural balance \cite{Cartwright} (also termed as the Heider balance \cite{Heider}) has been attracting attention of physicists for at least the last fifteen years.
The considered topologies include triangular lattices \cite{2005.11402}, complete graphs \cite{Antal_2005,Kulakowski2005,Marvel,PhysRevE.99.062302,1911.13048,PhysRevE.100.022303,2009.10136}, and complex networks \cite{Gawronski_2005a,PhysRevE.68.036122,PhysRevE.90.042802,Gorski_2017}; for review, see Refs.~\onlinecite{Kulakowski_2007,Krawczyk}. 
In all those cases, the applied models describe dynamics of negative or positive links representing hostile or friendly attitudes among actors decorating nodes of an underlying network.
The four possible arrangements of actors' attitudes in a single triad are presented in \Cref{fig:triads}.
The triads shown in \Cref{fig:triads}(a,c) %\Cref{fig:1a,fig:1c} 
are balanced in Heider's sense as they obey the following rules:
\begin{itemize}
\item a friend of my friend is my friend,
\item a friend of my enemy is my enemy,
\item an enemy of my friend is my enemy,
\item an enemy of my enemy is my friend.
\end{itemize}
%% ===============================================================
%% ===============================================================
The configurations presented in \Cref{fig:triads}(b,d) %\Cref{fig:1b,fig:1d} 
are imbalanced (i.e., they do not obey the rules given above) which results in actors' feelings of discomfort known as cognitive dissonance \cite{Festinger}.

\begin{figure}[htbp]
\includegraphics[width=\columnwidth]{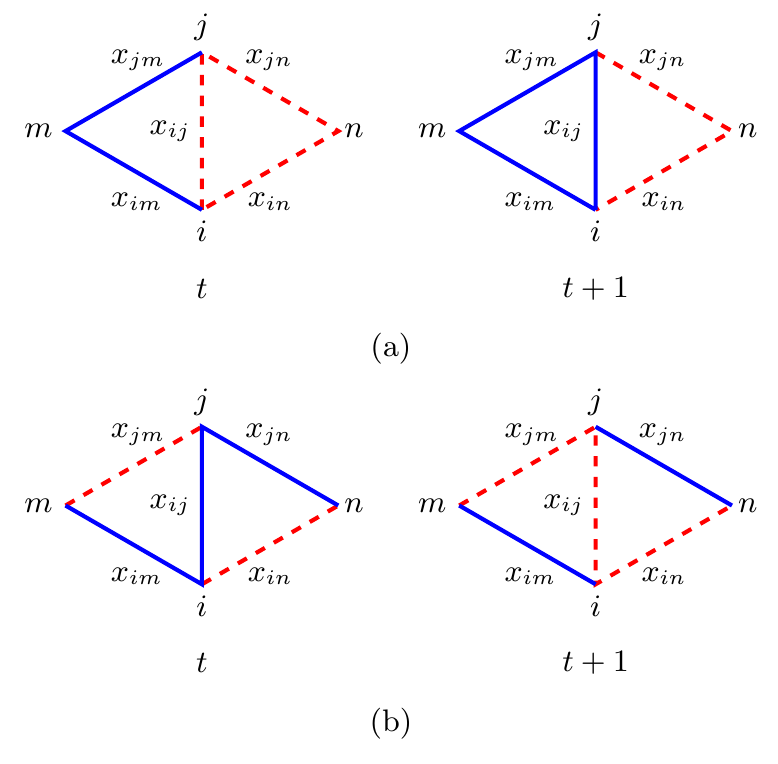}
\caption{\label{fig:rule}Examples of configuration of signed links $x_{im}$, $x_{in}$, $x_{jm}$, $x_{jn}$ which influence the link $x_{ij}$ in the next time step [(a) $x_{ij}(t)=-1 \to x_{ij}(t+1)=+1$, and (b) $x_{ij}(t)=+1 \to x_{ij}(t+1)=-1$] according to \Cref{eq:rule} in the deterministic case, i.e. for $T=0$.}
\end{figure}

In order to relieve this tension actors should change their attitudes to others by switching unfriendly or amicable relations into opposite ones.
Such process may be realized for two connected triads using the links dynamics given by
%% ---------------------------------------------------------------
\begin{equation}
\label{eq:rule}
x_{ij}(t+1)=\sign\big[x_{im}(t)x_{jm}(t)+x_{in}(t)x_{jn}(t)\big],
\end{equation}
%% ---------------------------------------------------------------
where $x_{ab}$ symbolizes friendly ($x_{ab}=+1$) or hostile ($x_{ab}=-1$) relation among the actors $a$ and $b$ (see \Cref{fig:rule} for two examples of system evolution towards the Heider balance, from an imbalanced state at time $t$ to a balanced state at $t+1$ after a single link flip $x_{ij}(t+1)=-x_{ij}(t)$).
%% ===============================================================
%% ===============================================================

We note that for $N$ actors placed in nodes of a regular triangular lattice, if all $2N$ triads are balanced, 
as presented in \Cref{fig:triads}(a,c),   % (\Cref{fig:1a,fig:1c})
then the system energy defined as
%% ---------------------------------------------------------------
\begin{equation}
\label{eq:U}
%U=-\dfrac{\sum_{i,j,k}x_{ij}x_{jk}x_{ki}}{2L^2} %% [km] {N} %% {3L^2}
U=-\dfrac{\sum_{i,j,k}x_{ij}x_{jk}x_{ki}}{2N}
\end{equation}
%% ---------------------------------------------------------------
is exactly equal to $U=-1$, which allows easy detection of the system balance without the triad-by-triad inspection.

Very recently, the deterministic evolution according to \Cref{eq:rule} has been enriched by introduction of the thermal noise simulated by the Glauber \cite{PhysRevE.100.022303} or the heat-bath \cite{PhysRevE.99.062302,1911.13048,2009.10136} dynamics.

In the latter case, the first order phase transition from (at least partially) balanced (with $-1\le U \le U^* <0$ \cite{PhysRevE.99.062302}) and ordered ($+1\ge \langle x_{ij}\rangle \ge x^*>0$ \cite{1911.13048}) state to an imbalanced (with energy $U=0$ \cite{PhysRevE.99.062302}) and disordered (with spatially averaged value of link strengths $\langle x_{ij}\rangle=0$ \cite{1911.13048}) state has been observed for a complete graph.

The transition takes place at the critical temperature $T^*$: for $T<T^*$ an ordered and balanced system state is observed, while for $T>T^*$ the time evolution drives the system to an imbalanced and disordered state.

In this paper we show that maintaining the thermal noise, but using the triangular lattice topology instead of a complete graph completely eliminates the ordered and balanced phases in the system.

%% ###############################################################
\section{\label{S:methods}Methods}
%% ###############################################################

The non-deterministic version of \Cref{eq:rule} reflecting the presence of thermal noise and using the heat-bath method \cite{Binder_1997} may be written as 
%% ---------------------------------------------------------------
\begin{subequations}
\label{eq:HB}
\begin{equation}
x_{ij}(t+1)=
    \begin{cases}
    +1 & \text{ with probability }p_{ij}(t),\\
    -1 & \text{ with probability }[1-p_{ij}(t)],
    \end{cases}
\end{equation}
where
\begin{equation}
    p_{ij}(t)=\frac{\exp[\xi_{ij}(t)/T]}{\exp[\xi_{ij}(t)/T]+\exp[-\xi_{ij}(t)/T]}, \label{eq:HBb}
\end{equation}
and
\begin{equation}
    \xi_{ij}(t)=x_{im}(t)x_{jm}(t)+x_{in}(t)x_{jn}(t),
\end{equation}
\end{subequations}
while $m$ and $n$ are the common neighbours of $i$ and $j$ nodes (see Fig.~\ref{fig:rule}), and $T$ is the temperature.
This way, the system is subject to a stochastic evolution.
Unlike when governed only by the tendency to minimize the number of imbalanced triads (in such case ultimately $U\to -1$), it is driven towards the thermal equilibrium corresponding to $U>-1$.
For example, starting with the state of paradise (with only friendly attitudes among neighbours), the energy monotonically increases from its minimal value, since the evolution rules allow for increase of energy with some non-zero probability.
The same is true for the standard Ising model when the starting point of evolution is a fully ferromagnetic ordered state and spin values evolution (governed by Metropolis, Glauber, or heat-bath dynamics: with the local rule for spin updates also minimizing the local spin energy) leads to a partially ordered or disordered state for $T>0$.

We apply \Cref{eq:HB} to find the time-evolution of $L=3N$ links on a triangular lattice of $N$ nodes assuming periodic boundary conditions.
A single Monte Carlo time step consists of $3N$ synchronously performed attempts to modify links in the system, i.e., each $x_{ij}$ link is subject to one such attempt.

%% ###############################################################
\section{\label{S:results}Results}
%% ###############################################################

%% ===============================================================
\begin{figure}
%\psfrag{U x}[][c]{$U$ \qquad\qquad $\langle x_{ij}\rangle$}
%% \psfrag{x}{$\langle x_{ij}\rangle$}
%% \psfrag{U}{$U$}
%% \psfrag{t}{$t$}
%% \psfrag{T=}{$T=$}
%% \includegraphics[width=\columnwidth]{Ux-t}
\includegraphics[width=\columnwidth]{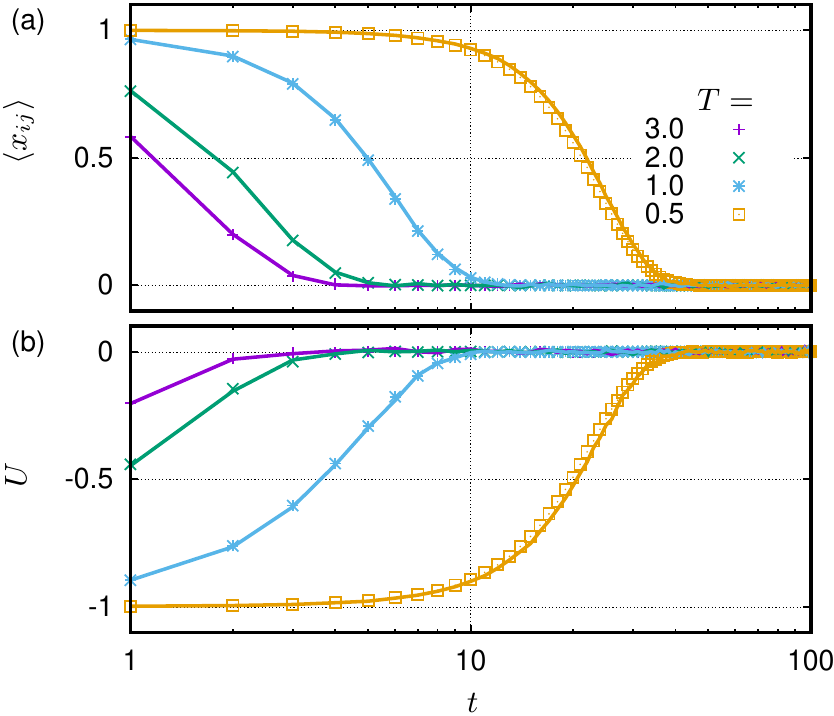}
\caption{\label{fig:xvst}Time evolution of (a) the average value of link strengths $\langle x_{ij}\rangle$, and (b) energy $U$ for various values of the social temperature $T$  
calculated for $N=400$ (lines) and $N=10000$ (symbols) nodes.
Initially, all the links are set to $+1$. 
The results are averaged over one hundred simulations.
The error bars are smaller than size of the symbols.}
\end{figure} 
%% ===============================================================

In \Cref{fig:xvst} we present the time evolution of the average value of link strengths $\langle x_{ij}\rangle$ and the system energy $U$ for various values of the social temperature $T$ and different numbers of nodes $N$.
The angle brackets $\langle\cdots\rangle$ indicate the averaging procedure performed over all $L$ links,
and then over one hundred simulations.
Initially, the Heider paradise is assumed, i.e., at $t=0$ each link is set to $x_{ij}=+1$.
Neither the system size $N$, nor the assumed temperature $T$ prevents reaching the unbalanced and disordered state with $\langle x_{ij}\rangle=0$ and $U=0$.

\begin{figure}[htbp]
% \psfrag{tau}[][c]{$\tau_0$ \qquad\qquad\qquad $\tau$}
%% \psfrag{tau}{$\tau$}
%% \psfrag{tau0}{$\tau_0$}
%% \psfrag{T}{$T$}
%% \psfrag{L=}{$L=$}
%% \includegraphics[width=\columnwidth]{tFirstFlip-t}
\includegraphics[width=\columnwidth]{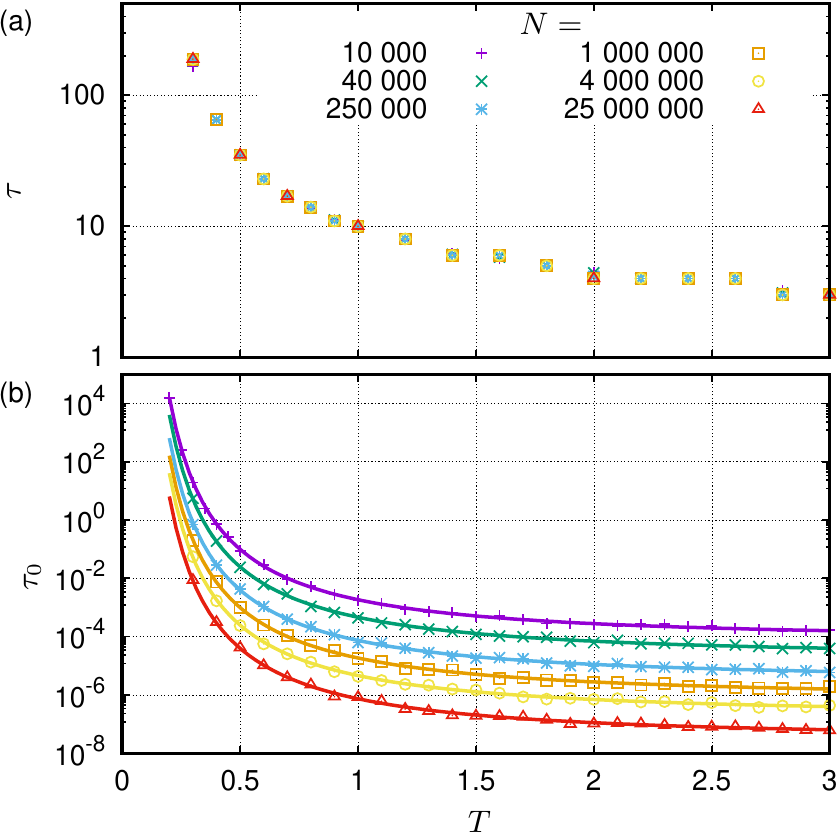}
\caption{\label{fig:tFirstFlip-t}%
(a) Time $\tau$ of reaching the stationary state and (b) time $\tau_0$ of the first link switching from $x_{ij}=+1$ to $x_{ij}=-1$.
The results are averaged over $R=100$ simulations. The error bars are smaller than size of the symbols and of the order of $1/\sqrt{R}$. Lines indicate the theoretical dependence $\tau_0=[1+\exp(4/T)]/(3N)$.}
\end{figure}

However, the time between the start of simulation and the first flip of any link value from $+1$ to $-1$
is equal to the reciprocal of probability that one link breaks out of the paradise state and changes the energy by $\Delta U = 4$.
Such probability is proportional to the number of links as the changes may happen independently and at any place, which means that the time of the first flip is given by
\begin{equation}
    \label{eq:tau0}
    \tau_0=\dfrac{1+\exp(4/T)}{3N} %{L^2}
\end{equation}
and tends to infinity in the limit of very low temperatures, $T\to 0^+$,
as shown in \Cref{fig:tFirstFlip-t} where both the above dependence (lines) and the results of simulation (points) are presented.
Thus for $T=0.1$ and $N=10^4$ no link switching from $+1$ to $-1$ is observed up to $t=10^8$ (i.e. until $Nt= 3\cdot 10^{12}$ trials).

\begin{figure}
%% \psfrag{x}{$\langle x_{ij}\rangle$}
%% \psfrag{U}{$U$}
%% \psfrag{t}{$t$}
%% \psfrag{T=}{$T=$}
%% \includegraphics[width=0.99\columnwidth]{xav-t-p095}
\includegraphics[width=\columnwidth]{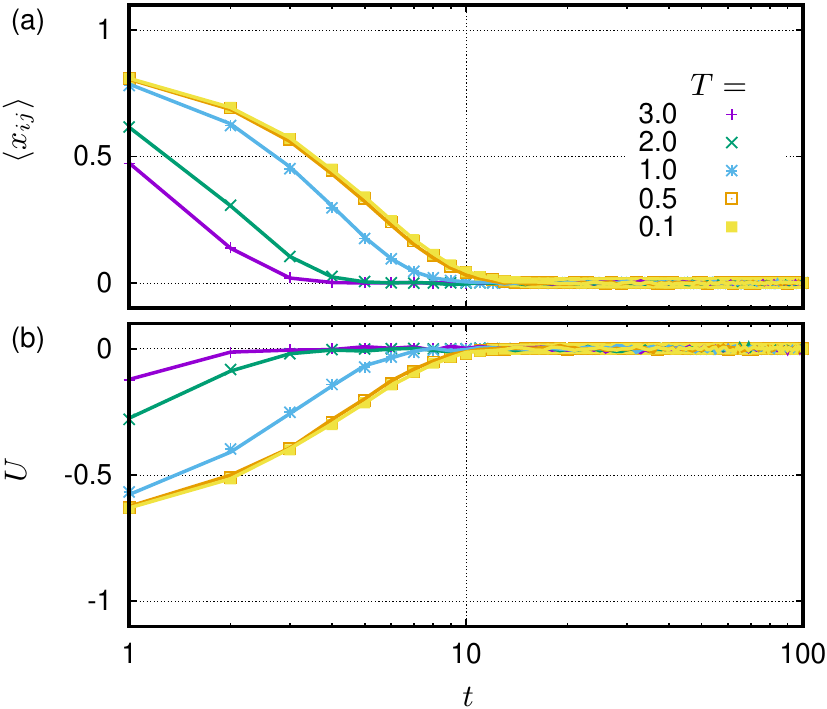}
\caption{\label{fig:xvst_95}Time evolution of (a) the average value of link strengths $\langle x_{ij}\rangle$ and (b) energy $U$ for $N=400$ (lines) and $N=10000$ (symbols) nodes and
for various values of social temperate $T$.
Initially, 95\% link values are set to $+1$,
%($\langle x_{ij}(t=0)\rangle\approx 0.9$).
hence $\langle x_{ij}(t=0)\rangle = 0.9$.
The results are averaged over one hundred simulations.
The error bars are smaller than the sizes of symbols.
}
\end{figure}

Further analysis of the system evolution in the low temperature limit is possible with a different initial state. 
For a starting point where only 5\% of the links are initially set to $x_{ij}=-1$ also systems kept in low temperature, $T=0.1$, reach unbalanced ($U=0$) and disordered ($\langle x_{ij}\rangle=0$) state, as demonstrated in \Cref{fig:xvst_95}.  

%% ===============================================================
%% ===============================================================

We assume that the system reaches the stationary state when $U$ and $|\langle x_{ij}\rangle|$ are smaller than $\varepsilon=10^{-2}$.
The time $\tau$ needed to reach the stationary state increases with decreasing temperature $T$, independently on the assumed system size $N$ (see upper part of \Cref{fig:tFirstFlip-t}). 
Deviations from this statement are observed only for small systems ($N<2000$) and only in low temperatures ($T<0.5$).

%% ===============================================================
%% ===============================================================

\begin{figure}
%% \psfrag{f}{$f$}
%% \psfrag{t}{$t$}
%% \includegraphics[width=0.99\columnwidth]{proc_triads_L100_T05_R100}
\includegraphics[width=\columnwidth]{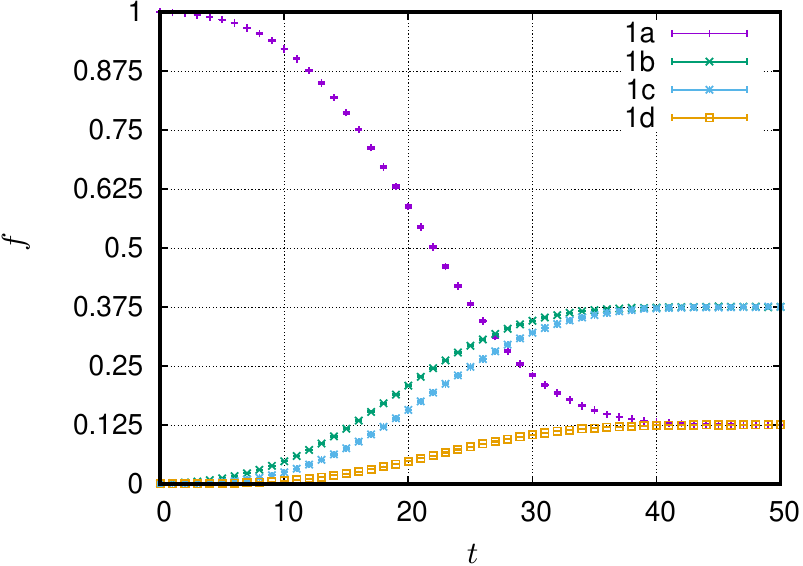}
\caption{\label{fig:triads-distribution}Time evolution of the fractions $f$ of triads presented in \Cref{fig:triads} %\Cref{fig:1a,fig:1b,fig:1c,fig:1d}
for $N=10^4$ and $T=0.5$. The results are averaged over one hundred simulations.
The error bars are smaller than the sizes of symbols.}
\end{figure}

To confirm the system transition to the disordered phase, let us take a closer look at the time evolution (see \Cref{fig:triads-distribution}) and the final probability distribution of triads presented in \Cref{fig:triads} for $N=10^4$ and $T=0.5$.
The obtained probabilities are $\frac{1}{8}$, $\frac{3}{8}$, $\frac{3}{8}$, and $\frac{1}{8}$ for the triads presented in %\Cref{fig:1a,fig:1b,fig:1c,fig:1d}, 
\Cref{fig:triads}(a), (b), (c), and (d),
respectively.
Different system sizes and temperatures give the same results, and we will discuss this issue shortly in Conclusions below.

%% ===============================================================
%% ===============================================================

%% ###############################################################
\section{\label{S:disc}Conclusions}
%% ###############################################################

In contrast to the stochastic evolution of the system with hostile and friendly attitudes among actors on a complete graph \cite{1911.13048} we show that {\em the triangular lattice topology successfully prevents the occurrence of even partial Heider balance}.

Starting at the state of paradise ($\forall i,j: x_{ij}(t=0)=+1$) it is just a matter of time $\tau$, when the thermally driven evolution (governed by the heat-bath algorithm) of the system leads to an unbalanced and disordered state. 
Contrary to the previously obtained results \cite{Antal-2006130,ISI:000231060800016} {\em the relaxation time $\tau$ does not depend on the system size}.

The probabilities distribution of various triads presented in \Cref{fig:triads}(a-d) %\Cref{fig:1a,fig:1b,fig:1c,fig:1d}
changes from $(1,0,0,0)$ at $t=0$ to $\left(\frac{1}{8},\frac{3}{8},\frac{3}{8},\frac{1}{8}\right)$ at $t\to\infty$.
The latter is in agreement with the probabilities 
\[ \binom{3}{k}p^k(1-p)^{3-k} \]
of $k=3$, 2, 1, and 0 successes in three Bernoulli trials, when the probability of a success is equal to $p=\frac{1}{2}$.
The result accentuates the randomness of the final link values distribution and emphasizes the system disorder for any positive temperature. 

What is minimized for $T>0$ is the free energy, $U-TS$, where $S$ is the entropy of the system.
Therefore, whether the ordered or the disordered phase is preferred depends on the relation between $U$ and $S$.
For example, energy $U$ in a complete graph is proportional to $N^3$, while the number of links $L$ is proportional to $N^2$.
Hence the number of states of links is $2^{N^2}$, and the entropy is $N^2\cdot\ln{2}$.
The energy prevails and the critical temperature increases to infinity in the thermodynamic limit (see Ref.~\onlinecite{1911.13048}).
On the other hand, in the triangular lattice both $U$ and $L$ are proportional to $N$, hence $S$ is proportional to $\ln{2^L}=L \ln{2}$ which is linear with $N$ as well as energy $U\propto N$, and allows for a finite critical temperature.
In particular, in the case of the Watts--Strogatz network with large neighborhood the critical temperature remains finite~\cite{2008.06362}.
However, for a small range of interaction  the critical temperature calculated numerically is approximately  zero (note that in the Watts--Strogatz network both $U$ and $L$ are proportional to $N$).
Our results indicate that in this aspect the triangular network is similar to  the classical unrewired Watts--Strogatz network~\cite{Watts_1998}. 
In both cases the critical temperature is close to zero within the numerically accessible accuracy.
We also note that in the case of parallel dynamics (cellular automaton approach) for the triangular lattice even at $T=0$ the energy $U$ does not reach the minimal value of $-1$.  
This result is confirmed also by the mean field calculations~\cite{2005.11402}
and suggests that indeed in the triangular lattice the density of energy is rather small, and therefore the critical temperature is close to zero.

In summary, we show that introducing noise to the system of hostile and friendly attitudes on triangular lattice leads to the disordered and imbalanced state independently on the system size and for any positive temperature.
In other words, the critical temperature of the system approaches zero. 
None of the earlier studies devoted to the problem of structural balance, and conducted for various topologies and different schemes of updating the link values, have revealed such complete lack of any signs of
the balanced state.

Our results indicate that at least in certain cases the behavior of the system does not have to follow what was so far believed to be a general tendency towards a global structural balance. 
Signatures of such possibility have been already provided, for instance in studies of bilayer networks \cite{Gorski_2017}.

\section*{Acknowledgements}
We are grateful to Krzysztof Ku{\l}akowski for fruitful discussions and comments on the manuscript.

\section*{Data Availability Statement}
The data that support the findings of this study are available from the corresponding author upon reasonable request.

%% \bibliography{heider,km}

\end{document}